\documentclass[aps,pre,groupedaddress,twocolumn,floatfix,showpacs]{revtex4}
\usepackage[dvips]{graphicx}
\usepackage{amssymb}
\usepackage{amsmath}

\begin{document}
\title{Signature of ray chaos in quasibound wave functions for a
  stadium-shaped dielectric cavity}

\author{Susumu Shinohara}
\author{Takahisa Harayama}

\affiliation{Department of Nonlinear Science, ATR Wave Engineering
Laboratories, 2-2-2 Hikaridai, Seika-cho, Soraku-gun, Kyoto, 619-0288,
Japan}

\begin{abstract}
Light emission from a dielectric cavity with a stadium shape is
studied in both ray and wave models. For a passive cavity mode with
low loss, a remarkable correspondence is found between the phase space
representation of a quasibound wave function and its counterpart
distribution in the ray model.  This result provides additional and
more direct evidence for good ray-wave correspondence in low-loss
modes previously observed at the level of far-field emission pattern
comparisons.
\end{abstract}


\maketitle

Directional lasing emission is one of the most highlighted features of
two-dimensional microcavity lasers \cite{ARC}.
In interpreting the appearance of emission directionality and its
dependence on cavity shape, a ray dynamical model has proven useful
\cite{ARC, Hentschel01, Schwefel04, Fukushima04, SB.Lee06,
Lebental06}.
In the standard version of the ray model, Frenel's law is applied to
describe the light emission process from a cavity without its
application being fully justified; Frenel's law is usually derived
when a plane wave is scattered at a planer dielectric interface.
For a cavity shape obeying integrable ray dynamics, one can
approximately make a connection between the ray picture based on
Frenel's law and wave solutions in the short-wavelength limit by using
the Eikonal method \cite{Tureci03}.
Besides, even for a nonintegrable cavity, one can associate its stable
ray trajectory (if it exists) to a class of wave solutions by the
Gaussian-optical method \cite{Tureci02}.
For a fully chaotic cavity, however, one lacks a method to relate ray
trajectories with wave solutions.
Whereas establishing ray-wave (or classical-quantum) correspondence in
closed chaotic systems has been very matured in the field of quantum
chaos \cite{QC}, it is still an ongoing issue to make such a
correspondence in ``open'' systems \cite{Open_sys}, one of which being
dielectric microcavities.

\begin{figure}[b]
\includegraphics[width=75mm]{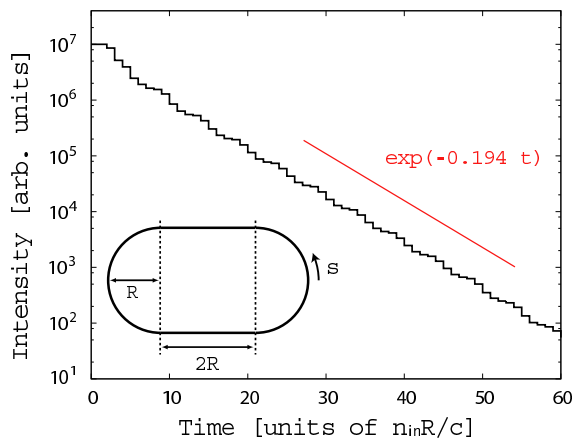}
\caption{(Color online) Exponential decay of light intensity in the
  ray model, where the time is measured in the unit of
  $n_{in}R/c$. Inset shows geometry of the stadium-shaped cavity.}
\label{fig:exp_decay}
\end{figure}

In this paper, we present numerical evidence showing that for a fully
chaotic cavity, there is significant correspondence between ray
dynamics and solutions of the Helmholtz equation, although we
currently lack justification for applying the ray model to a fully
chaotic cavity.
We consider a stadium-shaped cavity as shown in the inset of
Fig. \ref{fig:exp_decay}, whose internal ray dynamics is known to
become fully chaotic \cite{Bunimovich}.
Stadium-shaped cavities have actually been fabricated using materials
such as semiconductors \cite{Fukushima04} and polymers
\cite{Lebental06}, and stable lasing has been experimentally confirmed
in both materials.
In particular, for polymer cavities (refractive index $n\approx 1.5$),
the ray model predicts highly directional light emission, which can be
associated with the unstable manifolds of a short periodic trajectory
of the stadium cavity \cite{Schwefel04, Shinohara06}.
This highly directional emission has been experimentally observed, and
systematic agreement between experimental far-field patterns and those
obtained from the ray model has been reported in
Ref. \cite{Lebental06}.
Moreover, in recent work, we employed a nonlinear lasing model based
on the Maxwell-Bloch equations \cite{SB} to numerically simulate the
lasing of polymer stadium cavities and successfully obtained a highly
directional far-field emission pattern that agrees with the ray
model's prediction \cite{Shinohara06}.
The analysis of the passive cavity modes relevant for lasing revealed
that each of the low-loss (or high-$Q$) modes exhibits the far-field
emission pattern closely corresponding to the ray model's results.
The present work provides more direct and clearer evidence for this
ray-wave correspondence by showing that the phase space representation
of wave functions reproduces the ray model's distribution formed by the
stretching and folding mechanisms of ray chaos.

As a method to relate a wave function with ray dynamics, the Husimi
phase space distribution is often used \cite{SB.Lee06, Tureci05,
Hentschel03, SY.Lee}.
To accord with the definition of the phase space for the ray model,
where only the collisions with the boundary with outgoing momentum are
taken into account, it is appropriate to decompose a wave function into
radially incoming and outgoing components and then project the latter
onto the phase space.
Such decomposition has been implemented by using the expansion in
terms of the Hankel functions \cite{Tureci05}, which is, however, only
suited for a cavity shape slightly deformed from a circle.
Hence, here we introduce a different phase space distribution that
can be formally related with the ray model's distribution and directly
calculated from the wave function and its normal derivative at the
boundary.

First, we introduce a ray model incorporating Frenel's law.
In what follows, we fix the refractive indices inside and outside the
cavity as $n_{in}=1.5$ and $n_{out}=1.0$, respectively, and restrict
our attention to transverse magnetic (TM) polarization.
Inside the cavity, we regard the dynamics of a ray as a point particle
that moves freely except for bounces at the cavity boundary satisfying
the law of reflection.
We assign a ray trajectory a variable $\varepsilon(t)$ representing
intensity at time $t$, where $t$ is measured by trajectory length in
real space.
Due to the collision with the boundary at time $t$, the ray intensity
changes as $\varepsilon(t^{+})={\cal R}\,\varepsilon(t^{-})$, where
$t^{-}$ and $t^{+}$ are the times just before and after the collision
and ${\cal R}$ is the Fresnel reflection coefficient for TM
polarization \cite{Hecht}:
${\cal R}=[\sin(\phi-\phi_t)/\sin(\phi+\phi_t)]^2,$
where $\phi$ and $\phi_t$ are incident and transmission angles related
by Snell's law $n_{in}\sin\phi=n_{out}\sin\phi_t$.
Since we do not consider any pumping effect, $\varepsilon(t)$ is a
monotonically decreasing function.

Ray dynamics can be reduced to a two-dimensional area-preserving
mapping on the phase space defined by the Birkhoff coordinates
$(s,p)$, where $s$ is the curvilinear coordinate along the cavity
boundary and $p=\sin\phi$ is the tangential momentum along the
boundary.
The intensity leakage at the cavity boundary creates an ``open
window'' in the momentum space: Whenever a ray trajectory comes into
region $|p|<p_c=n_{out}/n_{in}$, it loses intensity by amount ${\cal
T}\,e$, where ${\cal T}$ is the transmission coefficient, i.e., ${\cal
T}=1-{\cal R}$, that can be expressed by sole variable $p$.

We assume that initially rays are distributed uniformly over the phase
space having identical intensities.
To study the statistical properties of the ray model, we focus upon a
time-independent distribution $P(s,p)$ that describes intensity flux
at the cavity boundary.
The usefulness of studying this distribution has been demonstrated in
Refs. \cite{Shinohara06, SY.Lee, Ryu06}.
Below we define this distribution for the ray model and later derive
the corresponding distribution for the wave model.

We denote the light intensity inside the cavity as ${\cal E}(t)=\sum_j
\varepsilon_j(t)$, where the sum is taken over the ray ensembles.
Its time evolution can be written as
\begin{equation}
\frac{d{\cal E}}{dt}=-\int_{0}^{S}ds\int_{-p_c}^{p_c}dp\,
{\cal T}(p)\,{\cal F}(s,p,t),
\label{eq:decay_ray}
\end{equation}
where ${\cal F}(s,p,t)$ represents intensity flux at the cavity
boundary and $S$ the total boundary length.

\begin{figure}[t]
\includegraphics[width=70mm]{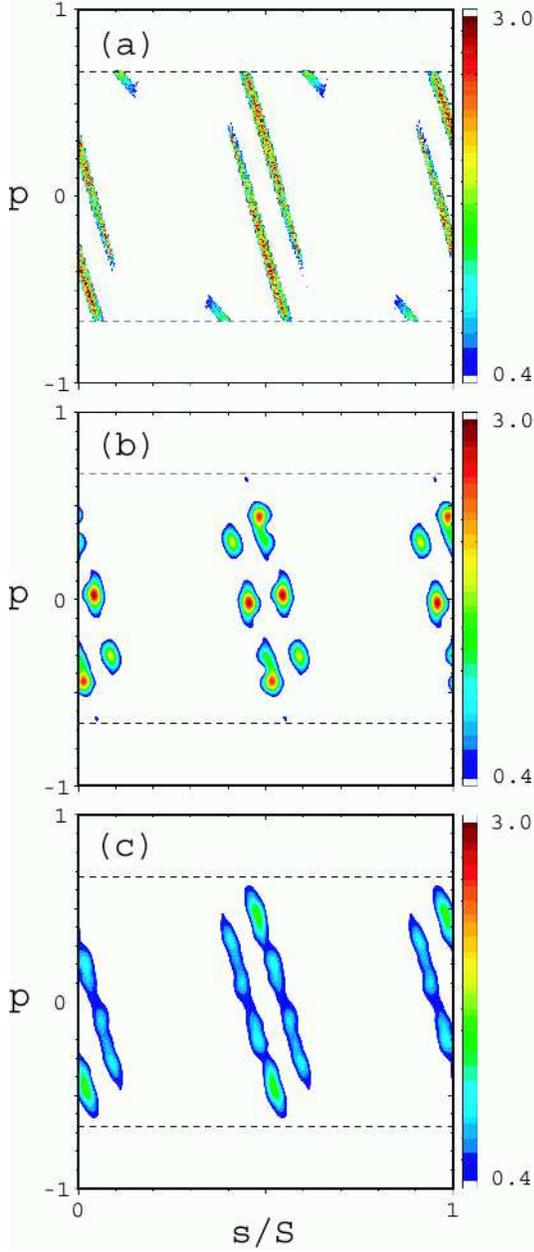}
\caption{(Color) (a) Intensity flux distribution $P(s,p)$ for the ray
  model. (b) Phase space distribution $H(s,p)$ of a wave function for a
  low-loss mode with $kR=100.00024-0.12667i$. (c) The average of
  $H(s,p)$ of the 21 lowest-loss modes. Dashed lines represent
  critical lines $p=\pm p_c$. Note that $P(s,p)$ and $H(s,p)$ are
  normalized to $\gamma_r$ as in Eq.(\ref{eq:exponent})}
\label{fig:phsp1}
\end{figure}

It has been numerically shown that ${\cal E}(t)$ exhibits exponential
decay behavior for stadium cavities \cite{Ryu06}.
Performing a numerical simulation with $10^7$ ray ensembles, we obtain
${\cal E}(t)$ as shown in Fig. \ref{fig:exp_decay}.
We can estimate the exponential decay rate as $\gamma_{r}\approx 0.194
\times c/(n_{in}R)$, where $c$ is the light speed
outside the cavity and $R$ the radius of the circular part of the
stadium cavity.
Exponential decay ${\cal E}(t)={\cal E}(0)\,e^{-\gamma_{r} t}$ can be
derived from Eq. (\ref{eq:decay_ray}) by assuming that ${\cal
F}(s,p,t)$ can be factorized as
%
${\cal F}(s,p,t) = F(s,p)\,{\cal E}(t)$ \cite{SY.Lee},
%
where the decay rate $\gamma_r$ can be expressed as
\begin{equation}
\gamma_{r}=\int_{0}^{S} ds \int_{-p_c}^{p_c} dp\,P(s,p).
\label{eq:exponent}
\end{equation}
Here, we put $P(s,p)={\cal T}(p) F(s,p)$ for convenience.
$P(s,p)$ describes how the rays' intensities are transmitted outside
the cavity and becomes important when trying to understand the
relation between emission directionality and the phase space
structures of ray dynamics \cite{Schwefel04, SY.Lee, Shinohara06}.
Figure \ref{fig:phsp1} (a) shows a numerically obtained distribution
$P(s,p)$.
As explained in detail in Refs. \cite{Schwefel04, Shinohara06}, the
structure of the high-intensity regions of $P(s,p)$ can be well fitted
by the unstable manifolds of a pair of unstable four-bounce periodic
trajectories; one is located just above critical line $p=p_c$ and the
other just below $p=-p_c$. 

\begin{figure}
\includegraphics[width=70mm]{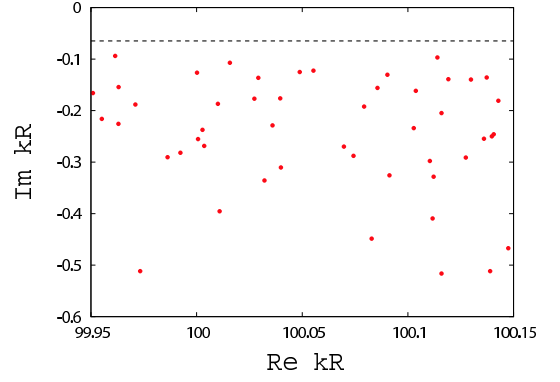}
\caption{(Color online) Distribution of resonances for $k_r R \approx
  100$. Prediction from the ray model is plotted in a dashed line.}
\label{fig:resonances}
\end{figure}

\begin{figure}
\includegraphics[width=70mm]{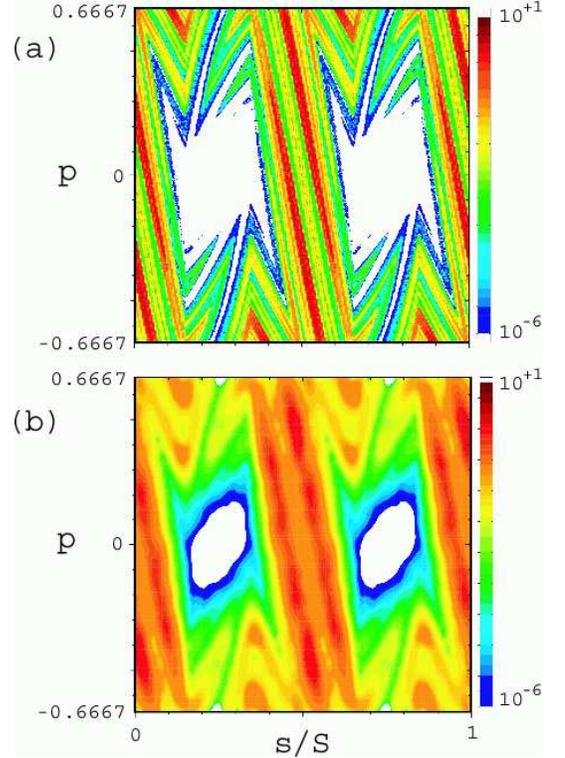}
\caption{(Color) (a) Log$_{10}$ plot of the intensity flux
  distribution $P(s,p)$ shown in Fig. \ref{fig:phsp1} (a). (b)
  Log$_{10}$ plot of the averaged distribution $H(s,p)$ shown in
  Fig. \ref{fig:phsp1} (c). Only region $|p|\leq p_c$ is shown.}
\label{fig:phsp2}
\end{figure}

Let us next treat the light field by the Maxwell equations.
For a two-dimensional cavity, the $z$-component of the TM electric
field is written as $E_{z}(x,y,t)=\mbox{Re}[\psi(x,y)\,e^{-i\omega
t}]$, where $\psi(x,y)$ satisfies the Helmholtz equation
$[\nabla_{xy}^2 + n^2(x,y)\,k^2]\,\psi=0$
and $\omega=c k$.
For a dielectric cavity, the eigensolutions of the Helmholtz equation
become quasibound states (or resonances) characterized by complex
wave numbers $k=k_r+ik_i$ with $k_i<0$.
Wave numbers $k$ and wave functions $\psi$ can be numerically obtained
by the boundary element method \cite{Wiersig03}.
In Fig. \ref{fig:resonances}, we plot the distribution of the
resonances for $k_r R\approx 100$.
For the wave description, the light intensity decay rate $\gamma_w$ is
written as $\gamma_w=2c|k_i|$.
Equating $\gamma_w$ with $\gamma_r$ evaluated in the ray simulation,
one obtains the ray model's estimate of the $k_i R$ value, i.e., $k_i
R=-0.194/(2n_{in})\approx -0.0647$, which turns out in this case to
give an upperbound of the $k_i R$ values as shown in
Fig. \ref{fig:resonances}.
It is an interesting problem to establish a precise correspondence
between $\gamma_r$ and $\gamma_w$ by a semiclassical argument, which
however we will not pursue here.

Next, we derive a distribution for the wave description that
corresponds to $P(s,p)$, formulating the intensity decay process as in
the ray model.
The light intensity of the cavity is written as ${\cal
E}=\int\int_{\cal D} dx dy \frac{1}{2}(\epsilon \vec{E}^2 + \mu
\vec{H}^2)$, where ${\cal D}$ represents the area of the cavity and
$\epsilon$ and $\mu$ are electric permittivity and magnetic
permeability, respectively.
The time evolution of ${\cal E}$ can be written as
\begin{equation}
\frac{d {\cal E}}{dt}=-\int_0^S ds\,{\cal S}(s,t),
\label{eq:decay_wave}
\end{equation}
where ${\cal S}(s,t)$ is the component of the Poynting vector normal
to the cavity boundary, i.e.,
${\cal S}(s,t)=c E_z (-\nu_x H_y + \nu_y H_x)$, 
where $\vec\nu$ is a unit vector normal to the cavity boundary.
In the TM case, $H_x$ and $H_y$ are determined from $E_z$ through
$\partial E_z/\partial y=-(\mu/c)(\partial H_x/\partial t)$ and
$\partial E_z/\partial x=(\mu/c)(\partial H_y/\partial t)$.
${\cal S}(s,t)$ contains terms rapidly oscillating in time with
frequency $2 c k_r$.
We smooth out this rapid oscillation by 
%
$\Bar{\cal S\,}(s,t)=\frac{1}{T}\int_t^{t+T}d\tau {\cal S}(s,\tau)$
%
with $T=2\pi/(ck_r)$.
Assuming $k_r\gg |k_i|$, which is valid in the low-loss and
short-wavelength limit, one obtains
\begin{equation} 
\Bar{\cal S\,}(s,t) = \frac{c e^{2c k_i t}}{2\mu k_r} \mbox{Im}
\left[
\psi^*(s) \partial_v \psi(s)
\right],
\label{eq:bar_S}
\end{equation}
where $\partial_{\nu}=\vec{\nu}\cdot\nabla$.
Moreover, we coarse-grain spatial variations smaller than the
wavelength by applying Gaussian smoothing as follows:
\begin{equation}
\Bar{\Bar{{\cal S}\,}}(s,t)=
\frac{1}{\sigma\sqrt{\pi}} \sum_{n=-\infty}^{\infty}
\int_{0}^{S} ds'\,e^{\left\{-\frac{(s'-s-nS)^2}{\sigma^2}\right\}}
\Bar{\cal S\,}(s',t),
\label{eq:barbar_S}
\end{equation}
where $\sigma=\sqrt{S/(2n_{in}k_rR)}$.
Plugging the right-hand side of Eq. (\ref{eq:bar_S}) into $\Bar{\cal
S\,}$ in Eq. (\ref{eq:barbar_S}), we obtain the following expression
for $\Bar{\Bar{{\cal S\,}}}(s,t)$:
\begin{equation}
\Bar{\Bar{{\cal S\,}}}(s,t)=\frac{c e^{2ck_i t}}{2\mu k_r}
\frac{1}{2\pi}
\int_{-\infty}^{\infty} dp\,H(s,p).
\label{eq:barbar_S_2}
\end{equation}
Here, $H(s,p)$ is a phase space representation of $\psi^*(s)\,
\partial_{\nu} \psi(s)$ similar to the Husimi distribution, defined by
\begin{equation}
H(s,p)=\mbox{Im}\left[h_{\psi}^*(s,p)\,
h_{\partial_{\nu}\psi}(s,p)\right], 
\end{equation}
where
\begin{equation}
h_f(s,p)=\int_{0}^{S} ds'\,G^*(s';s,p) f(s')
\end{equation}
and $G(s';s,p)$ is a coherent state for a one-dimensional periodic
system:
\begin{equation}
G(s';s,p)=\frac{1}{\sqrt{\sigma\sqrt{\pi}}}\sum_{n=-\infty}^{\infty}
e^{\left\{
-\frac{(s'-s-n S)^2}{2\sigma^2}+ip(s'-s-n S)
\right\}}.
\end{equation}
Comparing Eqs. (\ref{eq:decay_ray}) and (\ref{eq:decay_wave}) with
${\cal S}(s,t)$ being replaced with $\Bar{\Bar{{\cal S\,}}}(s,t)$, one
finds that $H(s,p)$ is the distribution that should be compared with
$P(s,p)$.

Calculating $H(s,p)$ for all the cavity modes shown in
Fig. \ref{fig:resonances}, we confirmed that for a low-loss mode,
$H(s,p)$ is predominantly supported on the high-intensity regions of
$P(s,p)$.
We show a typical example in Fig. \ref{fig:phsp1} (b), where we note
that to compare with $P(s,p)$ shown in Fig. \ref{fig:phsp1} (a), the
momentum is rescaled as $p/(n_{in}k_r R) \to p$ and $H(s,p)$ is
normalized to $\gamma_{r}$: $\int\int ds dp H(s,p)=\gamma_r$.
We plot the average of $H(s,p)$ of the 21 lowest-loss modes (i.e.,
those with $k_i R > -0.20$) in Fig. \ref{fig:phsp1} (c), which not
only proves that the localization on the high-intensity regions of
$P(s,p)$ is a common feature of low-loss modes, but also shows that
the correspondence with $P(s,p)$ becomes better by this averaging.
The correspondence between $P(s,p)$ and the averaged $H(s,p)$ can be
further revealed by plotting these distributions in logarithm scale as
shown in Fig. \ref{fig:phsp2}:
The log$_{10}$ plot of $P(s,p)$ reveals the structure of low-intensity
regions, which can be associated with the long-term behavior of the
unstable manifolds of the four-bounce periodic trajectories located
near the critical lines.
From Fig. \ref{fig:phsp2} (b), one can confirm that the averaged
$H(s,p)$ reproduces even the low-intensity regions of $P(s,p)$.

The ray-wave correspondence in low-loss modes provides a natural
explanation why experimental far-field patterns agree with the ray
model's prediction.
In experiments, lasing often occurs in multi-mode, so that a lasing
state can be considered as a ``superposition'' of multiple low-loss
modes.
The observation that better ray-wave correspondence is obtained after
the averaging over low-loss modes suggests that such a superposition
might enhance the ray-wave correspondence.

We thank M. Lebental for showing us unpublished data on ray model
simulations and S. Sunada for discussions.
The work at ATR was supported in part by the National Institute of
Information and Communications Technology of Japan.
\end{document}